\documentclass{article}

\PassOptionsToPackage{numbers}{natbib}
\usepackage[final]{neurips_2023_ml4ps}

\usepackage[utf8]{inputenc} 
\usepackage[T1]{fontenc}    
\usepackage[colorlinks=true, allcolors=blue]{hyperref} 
\usepackage{url}            
\usepackage{booktabs}       
\usepackage{amsfonts}       
\usepackage{nicefrac}       
\usepackage{microtype}      
\usepackage{xcolor}         
\usepackage{natbib} 
\usepackage{amsmath}
\usepackage{graphicx}

\graphicspath{{figs/}}

\newcommand*{\drm}{\mathrm{d}}
\newcommand*{\ve}[1]{\mathbf{#1}}

\providecommand{\sorthelp}[1]{} 

\title{Removing Dust from CMB Observations \\ with Diffusion Models}

\author{
  David Heurtel-Depeiges \thanks{Work done during an internship at the Flatiron Institute. $^\dagger$ Project supervisors.} \\
  Flatiron Institute\\
 Ecole Polytechnique, Institut Polytechnique de Paris\\
  \texttt{david.heurtel-depeiges@polytechnique.edu} \\
  \And
  Blakesley Burkhart \\
  Rutgers University \\
  Flatiron Institute \\
  \texttt{bburkhart@flatironinstitute.org} \\
  \And
  Ruben Ohana$^\dagger$ \\
  Flatiron Institute\\
  \texttt{rohana@flatironinstitute.org} \\
  \And 
  Bruno Régaldo-Saint Blancard$^\dagger$ \\
  Flatiron Institute\\
  \texttt{bregaldo@flatironinstitute.org} \\
}

\begin{document}

\maketitle

\begin{abstract}
    In cosmology, the quest for primordial $B$-modes in cosmic microwave background (CMB) observations has highlighted the critical need for a refined model of the Galactic dust foreground. We investigate diffusion-based modeling of the dust foreground and its interest for component separation.
    Under the assumption of a Gaussian CMB with known cosmology (or covariance matrix), we show that diffusion models can be trained on examples of dust emission maps such that their sampling process directly coincides with posterior sampling in the context of component separation. We illustrate this on simulated mixtures of dust emission and CMB. We show that common summary statistics (power spectrum, Minkowski functionals) of the components are well recovered by this process. We also introduce a model conditioned by the CMB cosmology that outperforms models trained using a single cosmology on component separation. Such a model will be used in future work for diffusion-based cosmological inference.
\end{abstract}

\section{Introduction}

The cosmic microwave background (CMB) is a key cosmological observable to constrain models describing the dynamical evolution of the Universe over its nearly 14 billion years of history~\citep{planck2016-l01}. However, CMB observations suffer from the contamination of various ``foreground'' signals (of astrophysical origin) and instrumental noise. For CMB analysis, these contaminants need to be removed, thus requiring precise component/source separation methods~(e.g., \cite{planck2016-l04, Jeffrey2021}). In particular, the active search for primordial $B$-modes in CMB polarization observations~\citep{Kamionkowski2016} has shed light on one of the main hurdles: the refined modeling of the thermal emission of interstellar dust grains radiating on top of the CMB, also known as the "dust foreground"~\citep{pb2015}.

Over the past decades, various models of the dust foreground have been developed (e.g., \cite{Allys2019, Finkbeiner1999, planck2013-p06b, Vansyngel2017,  Regaldo2020, Regaldo2021, Regaldo2023}). A recent series of works has notably leveraged the dazzling progresses of deep generative models for dust modeling~\citep{Aylor2021, Thorne2021, Krachmalnicoff2021}. In particular, \citep{harvardworkshop} made a first application of diffusion models for dust modeling and data denoising. We now take a new step in that direction by leveraging diffusion-based dust models for dust/CMB component separation.

Recently, diffusion models (DMs) have achieved state-of-the-art performance in generating high-quality natural images (e.g., ~\citep{ho2020denoising, Saharia2022, Ramesh2022, Rombach2022}). These models are trained on a denoising objective, with the goal to remove injected noise from images sampled from the target distribution. This training framework can be formalized in terms of diffusion processes that gradually add noise over time \citep{song2021scorebased}. In this work, we demonstrate that this formalism naturally applies to CMB component separation provided that the CMB can be assumed Gaussian.

We formalize our problem as follows. We observe an image $\ve{y} \in \mathbb{R}^{N\times N}$ that is a additive mixture of a dust signal $\ve{x}$ and CMB $\ve{\varepsilon}$ (with $\ve{x}, \ve{\varepsilon} \in \mathbb{R}^{N\times N}$):
\begin{equation}
    \ve{y} = \ve{x} + \ve{\varepsilon}.
\end{equation}
We assume that $\ve{\varepsilon} \sim \mathcal{N}(0, \sigma^2\ve{\Sigma}_\ve{\phi})$, where $\ve{\Sigma}_\ve{\phi}$ is the normalized covariance of the CMB parametrized by cosmological parameters $\ve{\phi}$, and $\sigma$ is the CMB amplitude. In that context, given a cosmology $\phi$ and a prior distribution on dust $p(\ve{x})$, the goal of component separation is to sample $p(\ve{x} | \ve{y}, \ve{\phi})$.\footnote{Note that, since $\ve{y} = \ve{x} + \ve{\varepsilon}$, a sample of $p(\ve{x} | \ve{y}, \ve{\phi})$ trivially yields a sample of $p(\ve{\varepsilon} | \ve{y}, \ve{\phi})$.}

In this paper, we show that, given a few hundred examples of dust maps, DMs can be trained so that their sampling process coincides with the sampling of $p(\ve{x} | \ve{y}, \ve{\phi})$. We also show that usual summary statistics (power spectrum, Minkowski functionals) are recovered by this process. 

\textit{Organization of the paper:} we describe our data in Sect.~\ref{sec:data} and present our method in Sect.~\ref{sec:method}. We then show our results in Sect.~\ref{sec:results} and discuss perspectives in Sect.~\ref{sec:conclusion}.

\section{Description of the Data}
\label{sec:data}

\begin{figure}
    \centering
    \includegraphics[width=0.8\hsize]{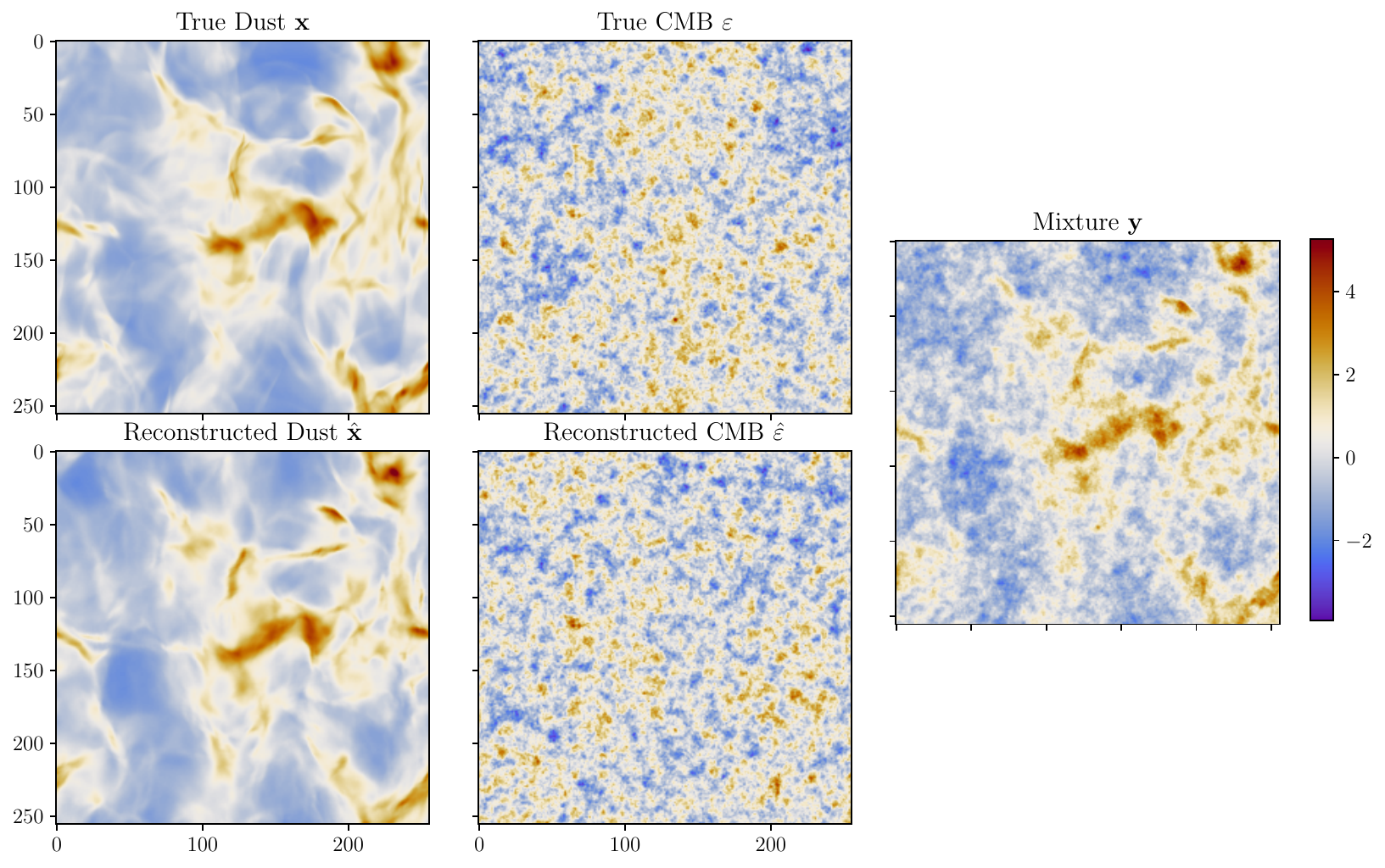}
    \caption{Observed and reconstructed maps. The true dust $\ve{x}$ and CMB $\ve{\varepsilon}$ maps compose the observed mixture $\ve{y}$. We reconstruct  the dust $\hat{\ve{x}}$ and CMB $\hat{\ve{\varepsilon}}$ with our diffusion model. The CMB maps are consistently rescaled for better visualization. The global unit is arbitrary.}
    \label{fig:imgs_data_soursep}
\end{figure}

\textbf{Dust Emission Maps.} We construct simulated dust emission maps in total intensity from a turbulent hydrodynamic simulation of the diffuse interstellar medium taken from the CATS database~\citep{burkhart2020catalogue}. We assume here that dust emission is proportional to the gas density of the simulation. Simulated maps then simply correspond to gas column density maps. We provide details on the hydrodynamic simulation in App.~\ref{app:simulation}.
Our dataset consists of 991 dust emission maps of size $256\times256$ (for an example, see Fig.~\ref{fig:imgs_data_soursep} top left panel).

\textbf{CMB maps.} If we neglect secondary anisotropies, CMB anisotropies are extremely well described by an isotropic Gaussian random field on the sphere~\citep{planck2016-l01}, entirely characterized by its covariance matrix (or power spectrum). As a consequence, we can obtain CMB maps by sampling Gaussian draws with  covariance $\ve{\Sigma}_\ve{\phi}$ parametrized by cosmological parameters $\ve{\phi}$ (for an example, see Fig.~\ref{fig:imgs_data_soursep} top middle panel). In this work, we consider a standard cosmological model. We only vary the cosmological parameters $H_0$ and $\omega_b$ and choose for the remaining ones fiducial values consistent with \textit{Planck} 2018 analysis~\citep{planck2016-l01}. We work with simulated maps covering an effective angular surface of approximately $34\times 34\,\mathrm{deg}^2$. We refer to App.~\ref{app:cmb_generation} for further technical details.

\section{Method}
\label{sec:method}

The contaminated CMB observation $\ve{y}$ can be viewed as the result of a forward diffusion process mapping the dust signal $\ve{x}$ to $\ve{y}$. We leverage DMs to learn the corresponding reverse diffusion process, thus enabling component separation.

\subsection{Diffusion Models: Formalism}

Given a forward diffusion process $(\ve{z}_t)_{t \in [0, 1]}$ mapping the target distribution $p_0$ to a simple Gaussian distribution $p_1$, DMs are trained to reverse this process (i.e. to map $p_1$ to $p_0$).
In this work, we consider the following forward diffusion process in $\mathbb{R}^d$ with $d=N^2$:
\begin{equation}
\label{eq:forward_sde}
    \drm\ve{z}_t = -\frac{1}{2}\beta(t)\ve{z}_t\,\drm t + \sqrt{\beta(t)}\ve{\Sigma}_\ve{\phi}^{\frac12}\,\drm \ve{w}_t,
\end{equation}
with $\beta : [0, 1] \rightarrow \mathbb{R}_+$ a scalar positive function, $\ve{\Sigma}_\ve{\phi}\in\mathbb{R}^{d\times d}$ the covariance matrix of the CMB, and $\ve{w}_t$ a standard $d$-dimensional Wiener process. We use a linearly increasing $\beta$ adjusted so that, approximately, $p_1 \sim \mathcal{N}(\ve{0}, \ve{\Sigma}_\ve{\phi})$.
The reverse process is also a diffusion process~\citep{song2021scorebased}:
\begin{equation}
\label{eq:backward_sde}
    \drm\ve{z}_t = \left[-\frac{1}{2}\beta(t)\ve{z}_t-\beta(t)\ve{\Sigma}_{\ve{\phi}}\nabla_{\ve{z}_t}\log p_t(\ve{z}_t)\right]\,\drm t + \sqrt{\beta(t)}\ve{\Sigma}_\ve{\phi}^{\frac12} \,\drm \overline{\ve{w}}_t,
\end{equation}
where $\overline{\ve{w}}_t$ denotes a standard backward-time Wiener process.
One can then generate samples of $p_0$ by solving this reverse SDE starting from $\ve{z}_1 \sim \mathcal{N}(\ve{0}, \ve{\Sigma}_\ve{\phi})$ (see App.~\ref{app:sampling} for further details).

\subsection{Posterior Sampling and Reverse Process}

We use the forward process in Eq.~\eqref{eq:forward_sde} to map the dust prior distribution $p_0 = p_{\rm dust}$ to the CMB distribution $p_1 = p_{{\rm CMB} | \mathbf{\phi}}$, and train a diffusion model on mixtures of dust and CMB maps to learn the inverse mapping. This enables sampling of dust maps conditioned on CMB seeds.

\looseness = -1
DMs are commonly trained using an identity covariance matrix, $\ve{\Sigma}_\ve{\phi} = \mathbf{I}_d$, resulting in a standard Gaussian distribution for $p_1$. By instead incorporating the true statistics of the CMB through $\ve{\Sigma}_\ve{\phi}$ in the SDE, our approach is tailored to extracting the dust component from the observed mixture.

Indeed, assuming cosmology $\ve{\phi}$ known, observation $\ve{y}$ can be viewed as a (rescaled) sample of $\ve{z}_{t^\star}$ with $t^\star \in [0, 1]$ resulting from the forward process of Eq.~\eqref{eq:forward_sde}. Sampling the posterior $p(\ve{x}| \ve{y}, \ve{\phi{}})$ can be achieved by solving the reverse SDE of Eq.~\eqref{eq:backward_sde}, starting at time $t^\star$ and initialized with $\ve{z}_{t^\star} = \ve{y}$~\citep[see][Sect. 5]{anderson1982reverse}. In this work, we assume that $t^\star$ is given\footnote{It is a closed form function of $\sigma$.}. Future work will address the problem of inferring it from the data.

\subsection{Training}

While we could have trained a DM based on Eqs.~\eqref{eq:forward_sde} and \eqref{eq:backward_sde} for a fixed cosmology $\ve{\phi}$, physical applications involve unknown CMB parameters. Our long-term goal being to infer $\ve{\phi}$ from observations~$\ve{y}$, we choose here to train a DM conditioned by the cosmology $\ve{\phi}$. As a proof of concept, we only consider a dependency on cosmological parameters $H_0$ and $\omega_b$.  With $\ve{\phi} = (H_0, \omega_b)$, we choose a broad prior $p(\ve{\phi}) \sim \mathcal{U}([50, 90]\times [0.0075, 0.0567])$.

Our DM is a three-level UNet with ResBlocks, trained by minimizing the denoising score-matching loss~\citep{song2021scorebased}. We encountered substantial training instabilities due to the large  ($\sim10^4$) condition number of $\ve{\Sigma}_\ve{\phi}$. By renormalizing the loss and model outputs (details in App.~\ref{app:training}), we were able to mitigate these issues. Additional training specifics are provided in the Appendix.

\section{Results}
\label{sec:results}

\subsection{Generative Modeling}

As a first validation of our diffusion model, we investigate in App.~\ref{app:gen_modeling} its performances for the generation of dust emission maps. This first analysis shows that our model generalizes very well beyond its training set by producing visually and quantitatively realistic maps.

\begin{figure}
    \centering
    \includegraphics[width=0.9\hsize]{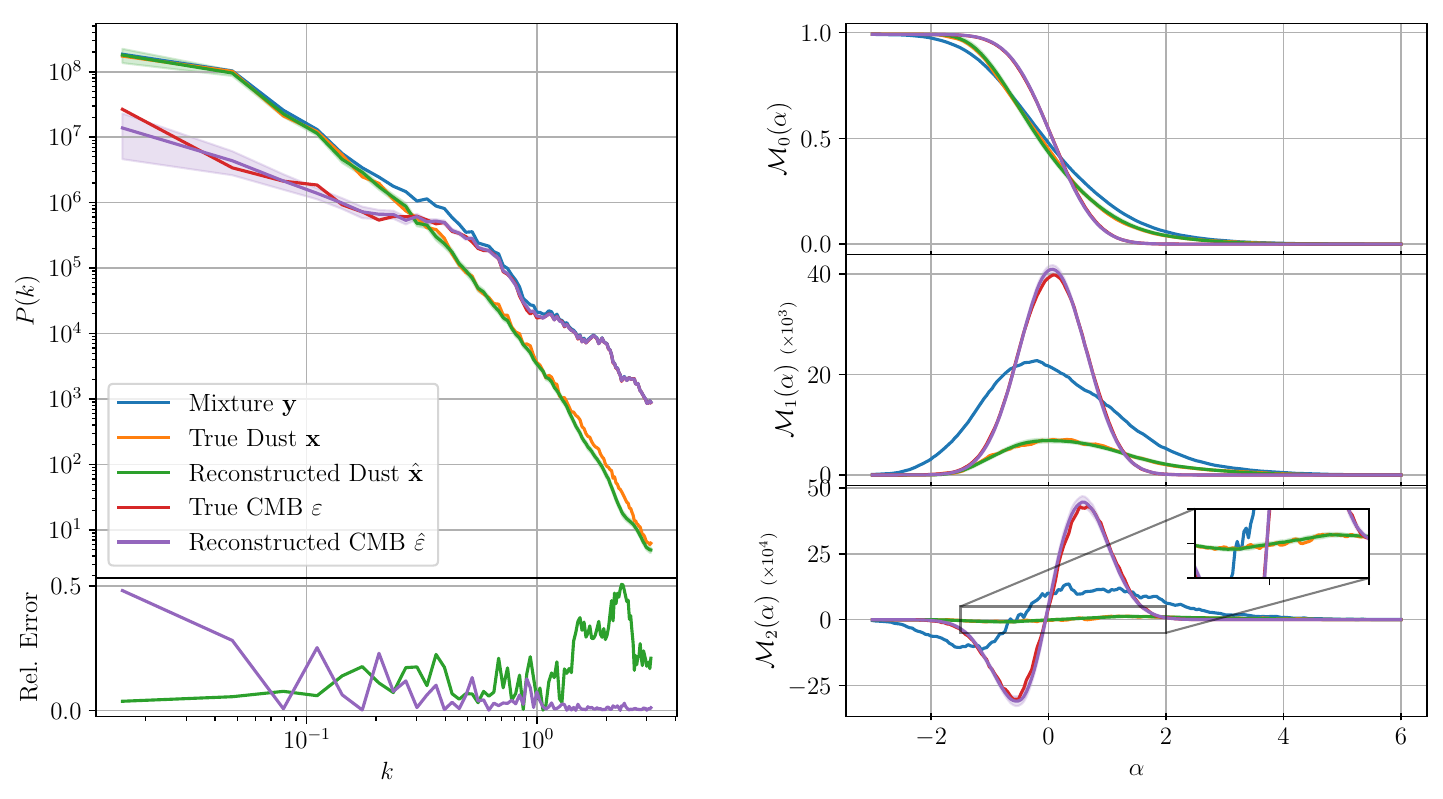}
    \caption{Reconstruction of summary statistics: power spectrum (left) and Minkowski functionals (right) of the mixture and the true and reconstructed dust and CMB maps.}
    \label{fig:compsep_stats}
\end{figure}

\subsection{Component Separation}

We evaluate our model's component separation performance on an observation $\mathbf{y}$ produced for a CMB map with $H_0=70$, $\omega_b=0.0321$ (i.e. center of prior range $p(\mathbf{\phi})$) and amplitude $\sigma = 0.6\,\sigma_\ve{x}$.

Fig. \ref{fig:imgs_data_soursep} shows the reconstructed CMB and dust maps. The large-scale features of $\hat{\mathbf{x}}$ clearly match those of $\mathbf{x}$. However, at small scales, where the CMB is dominant, we observe significant discrepancies. This was expected and underlies the probabilistic aspect of the reconstruction.

We also show in Fig. \ref{fig:compsep_stats} the reconstructed power spectra and Minkowski functionals. These match the statistics of the true components to a very good extent. The CMB spectrum is very well recovered, even at scales where the dust dominates. For dust, the power spectrum is overall well recovered. However, the relative error at small scales presents wavy patterns that exhibit a flaw of the reconstruction: these correspond to residual imprints of CMB power spectrum bumps that were not entirely smoothed out during reconstruction. Overall, the Minkowski functionals agree to a very good extent, underlying the success of our model in reconstructing non-Gaussian dust features.

We have  verified that the previous results are not specific to the choice of cosmology. However, we report slightly worse reconstructions for cosmologies at the edge of the prior range $p(\ve{\phi})$.

\subsection{Interest of a Cosmology-Dependent Model}

We have worked with a cosmology-dependent model mainly for (long-term) purposes of diffusion-based cosmological inference. Nevertheless, we observed that, while seemingly more intricate, it was not harder to train than a DM trained for a single cosmology. In particular, it did not require more training epochs. Moreover, our cosmology-dependent model proved to be on par or perform better than single-cosmology models. We have notably observed experimentally that it yields better power spectrum reconstruction for dust generative modeling. This suggests that our model has learned a more robust representation of dust, and illustrates the interest of multi-task learning for this problem.

\section{Conclusion and Perspectives}
\label{sec:conclusion}
In this work, we demonstrated that diffusion models can be naturally applied to separate cosmic microwave background (CMB) signals from dust contamination, under the assumption that the CMB follows a Gaussian distribution. We trained a model conditioned on cosmological parameters to map from a prior distribution of simulated dust maps to the known CMB distribution. We showed that such a model allows for posterior dust and CMB sampling given an observed mixture and assumed cosmology. In particular, we showed that key statistics of the individual components are accurately recovered through this process.
While based on simulations, this proof-of-concept establishes diffusion models as a promising approach for analyzing CMB observational data using arbitrary dust priors and less than a thousand dust examples. 
Future directions include developing a diffusion-based pipeline for inferring cosmological parameters $\ve{\phi}$ from contaminated CMB observations. Additional work is needed to extend this approach to polarization data and account for other contaminants like instrumental noise. Another challenge for the rightful application of our method to observational data will be to generalize this work to the case of weakly non-Gaussian CMB signal.

\begin{ack}

We would like to thank Fiona McCarthy for her help with the generation of the CMB simulations, as well as Jiequn Han, Ronan Legin, Charles Margossian, Chirag Modi,  Loucas Pillaud-Vivien, and Yuling Yao for valuable discussions.

\end{ack}


\bibliographystyle{abbrv}
\bibliography{bib/references.bib, bib/planck.bib}


\newpage
\appendix
\counterwithin{figure}{section}

\newpage

\section{Complements on Data}

\subsection{Hydrodynamic Simulation}
\label{app:simulation}

The simulation uses a third-order-accurate hybrid essentially non-oscillatory scheme~\citep{Cho2003} to solve the fluid hydrodynamic equations. It has periodic boundary conditions and an isothermal equation of state $p = c_{\rm s}^2 \rho$, with $c_{\rm s}$ the isothermal sound speed.
Turbulent motions are generated with a random large-scale solenoidal driving at a wave number $k\approx 2.5$ (i.e.~1/2.5 the box size) and the driving is continuous to prevent the decay of turbulence in one turn over time.
The simulation has $256^3$ resolution elements. Related simulations have been described and used in many previous works~\citep{Cho2003,kowal2007,2014ApJ...785L...1C,osti_22521545,bialy2017ApJ...843...92B,Portillo2018ApJ...862..119P}. They are part of the Catalog for Astrophysical Turbulence Simulations \cite{burkhart2020catalogue}.

The primary control parameters of the simulations are the dimensionless sonic Mach number, ${\mathcal{M}_{\rm s} \equiv  |\ve{v}|/c_{\rm s}}$, which in our case is ~7. $\ve{v}$ is the velocity, $c_{\rm s}$ and $v_A$ are respectively the isothermal sound speed and Alfvén speed.

\subsection{CMB Simulations}
\label{app:cmb_generation}

Given a set of cosmological parameters $\phi$, CMB power spectra are computed using \texttt{CAMB}~\citep{Lewis2000}. Gaussian draws on the sphere are then projected on $256\times 256$ patches using \texttt{pixell}~\footnote{\url{https://github.com/simonsobs/pixell}} with a pixel size of $8^\prime$. We only vary the cosmological parameters $H_0$ and $\omega_b$ in this work and use $\Omega_K = 0$, $\omega_c = 0.12$, $\tau = 0.0544$, $n_s = 0.9649$, ${\ln(10^{10} A_s) = 3.044}$, $m_\nu = 0.08$.

\section{Models and Training Details}
\label{app:training}
We withheld 10\% of the 991 images to validate the quality of generation/denoising of our diffusion model trained on the remaining 90\%.


\subsection{Architecture}\label{app:archi}

\textbf{Backbone Architecture.} 
Our score network is a three-level UNet with ResBlocks and a bottleneck of size $32\times 32$ or $16\times 16$. Each level has one ResBlock with three convolutions, whether descending or ascending and use GroupNorm as normalization with SiLU for activation. In each block, we rescale the skip-connections. Skip-connections across the bottleneck are concatenated and not added, contrary to internal skip-connections. We also enforce periodic boundary conditions by using 'circular' padding in our convolutions.

Time $t$ is also as seen an input, as in \citep{ho2020denoising,song2021scorebased}, and a sine/cosine time embedding is used. After being transformed by a MLP, this embedding is added at each block. (whose structure differs if the SDE is discretized a priori or if we train in continuous time) and an MLP-transformed version of this embedding is then added at the beggining of each block.

\begin{figure}
    \centering
    \includegraphics[width=\hsize]{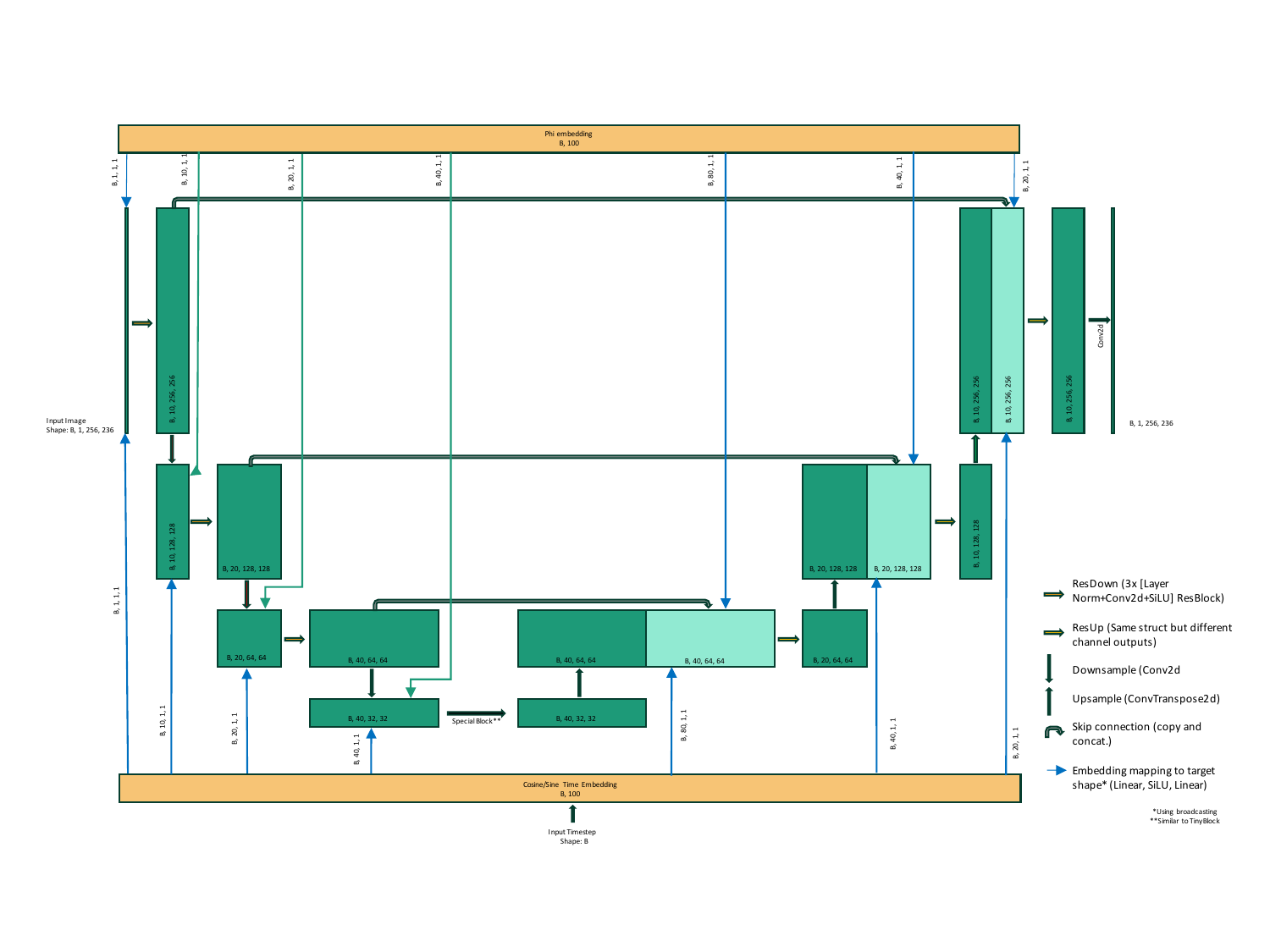}
    \caption{Architecture of our diffusion model for a parametrized family of SDE.}
    \label{fig:architecture}
\end{figure}

\textbf{Cosmology-Dependent Model.} 
In the case of a parametrized family of SDE (Eq.~\eqref{eq:forward_sde} and corresponding paragraph) we need to train a diffusion model to reverse a continuum of forward SDEs parametrized by $\phi$. We adapt the above architecture by adding a dependency to $\phi$ that is parallel to the dependency on time $t$. We embed $\phi$ with a single linear layer with activation and then, for each ResBlock, this initial embedding is further transformed by another MLP (see Fig.~\ref{fig:architecture}).

\subsection{Loss Functions}

\textbf{Single Cosmology.}
Our model is trained by minimizing the denoising score-matching loss which is:
\begin{equation}
\label{eqapp:singlenoise_loss}
     \mathcal{L}(\theta) = \mathbb{E}_t\left[~\lambda(t)\mathbb{E}_{\ve{z}_0,\,\ve{z}_t\vert\ve{z}_0}\left[\left\lVert \mathbf{s}_\theta(\ve{z}_t,t)+\mathbf{\Sigma}^{-1}\left(\ve{z}_t-\sqrt{\bar\alpha(t)}\ve{z}_0\right)/\left(1-\bar\alpha(t)\right)\right\lVert^2_2\right]\right],
\end{equation}
with $\bar\alpha(t)=\exp(-\int_0^t\beta(u)\mathrm{d}u)$. When generating samples, we then multiply the output of $\mathbf{s}_\theta$ by $\mathbf{\Sigma}$. This can induce numerical errors that are difficult to counteract if the matrix $\mathbf{\Sigma}$ is ill-conditioned (which is our case). Another downside is the fact that training seems extremely unstable and difficult to monitor with a loss that doesn't decrease. 

We found that re-normalizing what we learn could help dealing with both of issues: instead of $\mathbf{s}_\theta$, we learn $\mathbf{m}_\theta = - \sqrt{1-\bar\alpha(t)}\mathbf{\Sigma~s}_\theta$.

The square root part of this renormalization is standard, the $-1$ is for convenience but the $\ve{\Sigma}$ is new and its role is different than that of the square root. In DDPMs and score-based models, square root renormalization (or its equivalent in the case of different SDE) is implemented so that models try to learn a multivariate standard Gaussian instead of a quantity with widely varying variance.

\textbf{Multiple Cosmologies.}\label{app:MCMloss}
The generalized training loss is a straightforward extension of Eq.~\eqref{eqapp:singlenoise_loss}:
\begin{equation}\label{eq:multinoiseloss}
    \mathcal{L}(\theta)=\mathbb{E}_{t,\phi}\left\{~\lambda(t)\mathbb{E}_{\ve{z}_0,\,\ve{z}_t\vert(\ve{z}_0,\phi)}\left[\left\lVert \mathbf{m}_\theta(\ve{z}_t,t,\phi)-\frac{1}{\sqrt{1-\bar\alpha(t)}} \left(\ve{z}_t-\sqrt{\bar\alpha(t)}\ve{z}_0\right)\right\lVert^2_2\right]\right\}
\end{equation}

Training examples are generated by combining dust samples and CMB realizations for cosmological parameters randomly drawn from the prior $p(\phi)$. Since the computation of a single covariance matrix $\ve{\Sigma}$ takes a few seconds, we train neural emulator to approximate $\phi \rightarrow \ve{\Sigma}_\ve{\phi}$ as described in Appendix~\ref{app:emulator} for faster generation of training samples.

\subsection{Sampling Procedure}
\label{app:sampling}
The standard sampling scheme for an SDE is Euler-Murayama (equivalent to Euler's method for ODE). Given a SDE:
\begin{equation*}
	\mathrm{d}\ve{z}_t = \mathbf{f}(\ve{z}_t,t)\mathrm{d}t + \mathbf{G}(\ve{z}_t,t)\mathrm{d}\mathbf{w}_t,
\end{equation*}
and an initial point $\ve{z}_0$, Euler-Murayama sampling for a timestep $\Delta t$ follows this update rule:
\begin{equation*}
	\forall k,~~\ve{z}_{(k+1)\Delta t} = \ve{z}_{k\Delta t} + \Delta t\, \mathbf{f}(\ve{z}_{k\Delta t},k\Delta t) + \sqrt{\Delta t}~ \mathbf{G}(\ve{z}_{k\Delta t}, k\Delta t) \varepsilon,
\end{equation*}
with $\varepsilon\sim\mathcal{N}(0,\mathbf{I}_d)$.
Note that variants of Euler-Murayama exist with equivalent of the Runge-Kutta method or adaptive learning rate schemes. A very similar equation exists for sampling in reverse time.

For the majority of our experiments, we discretized SDEs before training and used therefore a fixed time schedule for SDE and ODE sampling. The rest of our implementation is close to that of \cite{song2021scorebased} with inputs from different freely available repository.

\subsection{Training heuristics}\label{subsecapp: training heuristics}

We performed a number of coarse grid search hyper-parameter optimization for different versions of the problem that we solve. In this section, we discuss failures on architectures tests.

\textbf{Overfitting.}
In comparison with diffusion models trained under white noise, our framework is much more prone to overfitting. This is made worse by the small dataset size we have had to work with. Our understanding is that because the noise (CMB) power spectrum value at small scales is much less than that of white noise, small and medium scales are less affected by our forward diffusion process, making it easier for the network to recognize structures in the images, therefore overfitting. 

We also tried our framework on choices of $\ve{\Sigma}$ not corresponding to CMB realizations. For some value of said covariance matrix that whose condition number is more extreme than that of CMB covariances, training was impossible because the model systematically overfitted.

For generating CMB realizations, we found there exists a range of model capacities that avoid overfitting. Specifically, UNet architectures with a large bottleneck size and a small number of channels tend to perform best.

\textbf{Choice of hyperparameters.} Apart from avoiding overfitting, the choice of hyper-parameters did not seem to impact much of model performance. We typically maxed out capacity without overfitting and models produced this way tended to have the best performance.

\textit{Learning rate.} We used AdamW with an inverse square root  schedule with a peak ranging from $4\times 10^{-4}$ to $10^{-3}$

\textit{Number of discretization steps.} Increasing this value in the SDE produced better results, as expected. Our understanding is that it is mainly due to small values of $t$ where details are sampled. At small scales, a small time step allows us to mitigate (in part) what can be observed in the power spectrum of samples from our generative models: oscillations in power spectrum values at small scales correspond to those of the CMB power spectrum (see Fig. \ref{fig:gen_stats}).

\textbf{Training.} When training our new diffusion models, we observed the minimum loss reached was substantially higher (1-2 orders of magnitude) than that of white noise diffusion models, depending on the settings and dataset. However, as with white noise models, we found it was essential to continue training long after the loss plateaued, occupying 80-90\% of total training time, since interrupting training too early resulted in poorer performance based on summary statistics and visual assessment.

\subsection{Overfitting Checks}\label{app:overfitting_checks}

In order to verify that the model is not overfitting, we check for mode collapse and dataset copying with a specific $L_2$ distance taking into account the periodic boundary conditions. This was particularly important as our models seem more prone to overfitting than model trained under white noise. Further discussion of why can be found in App.\ref{subsecapp: training heuristics}.

\subsection{Emulator for the map \texorpdfstring{$\ve{\phi} \rightarrow \mathbf{\Sigma}_\phi$}{phi -> S(phi)}}\label{app:emulator}

We train an emulator to compute efficiently the power spectrum of the CMB for arbitrary $\phi = (H_0, \omega_b)$ parameters withing a prior range. We use an architecture based on a simple MLP with two hidden layers and 100 hidden neurons per hidden layer.

\subsection{Hardware and training times.}
All of our training was performed on NVIDIA A100 GPUs. One epoch takes $\sim3$ seconds for approximately 1,000 images. We would usually train for between 10,000 and 20,000 epochs or between 10 and 20 hours (taking into account time for sampling during training). At inference, sampling 128 images with 1,000 steps in Euler-Maruyama takes $\sim 1$ minute. We found that for batch sizes larger than 16, sampling time is directly proportional to the number of images (as well as to the number of discretization steps).

\section{Model Performances for Dust Generative Modeling}
\label{app:gen_modeling}

\begin{figure}[!t!]
    \centering
    \includegraphics[width=\textwidth]{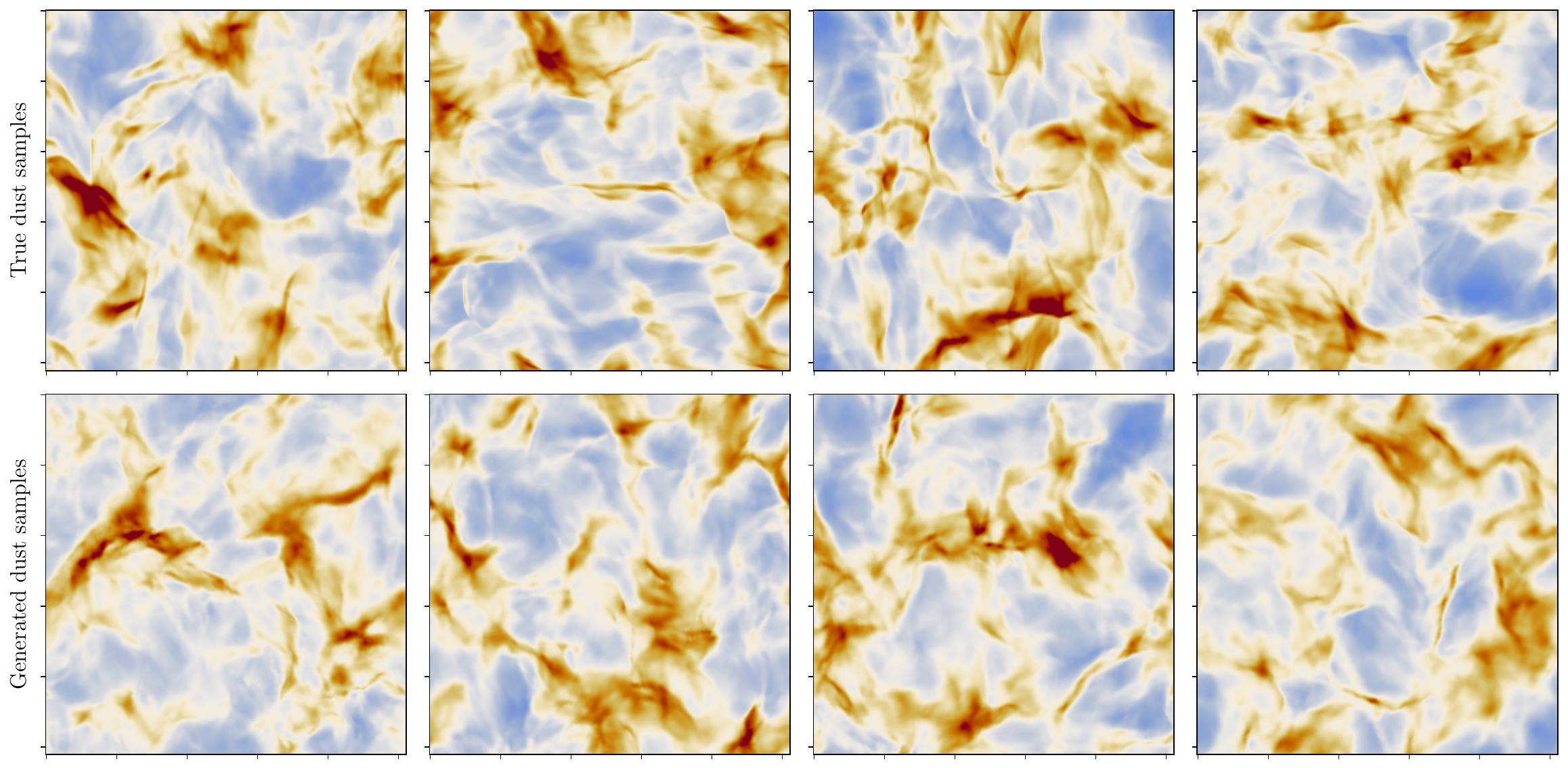}
    \caption{Examples of original (top row) and generated (bottom row) dust emission maps.}
    \label{fig:gen_samples_vs_original}
\end{figure}

\begin{figure}
    \centering
    \includegraphics[width=\hsize]{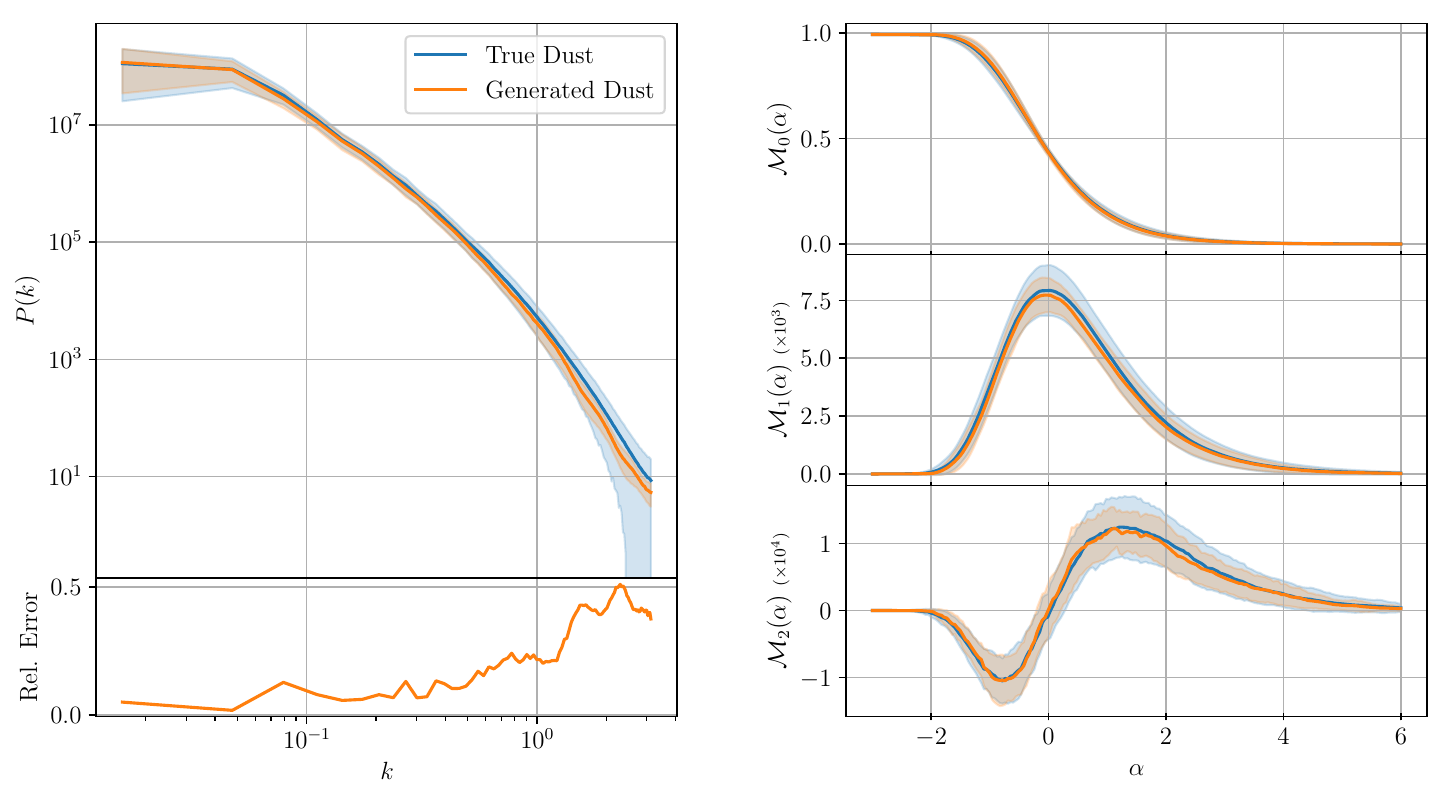}
    \caption{Comparison of the power spectrum statistics (left) and Minkowski functionals (right) between dust samples from our diffusion model and training dust data.}
    \label{fig:gen_stats}
\end{figure}

We assess the abilities of our model in terms of generative modeling by generating dust samples and comparing them to training examples using the following:

\textit{Visual assessment.} We compare in Fig.~\ref{fig:gen_samples_vs_original} examples of generated dust samples (bottom row) to a random selection of original samples (top row). Maps are visually indistinguishable, which is a demonstration of the quality of the learned generative model. We rule out any potential trivial overfitting concerns using a custom metric defined in ~\ref{app:overfitting_checks},
corroborating the ability of our model to generalize beyond its training set.

\textit{Quantitative assessment.} In Fig.~\ref{fig:gen_stats} we compare generated and original samples using the empirical power spectrum (left) and Minkowski functionals (right). The power spectrum captures Gaussian features, while Minkowski functionals quantify non-Gaussian information. Means and standard deviations are computed over 64 samples.

Overall, summary statistics match to a very good extent. Minkowski functionals are very well recovered both in terms of mean and standard deviation. The mean power spectrum is also well recovered especially at larger scales. However, we notice a small lack of power for the generated samples at small scales as well as lower standard deviations. We attribute this to three phenomena:  UNet averages at small scales, a lack of data, and a finite number of discretization steps as we approach $t=0$. We emphasize that for applications to source separation, errors made on the reconstruction of the dust power spectrum could directly impact the power spectrum of the reconstructed CMB. However, because these errors are mainly concentrated at small scales, where the CMB is dominant, the reconstruction error on the CMB power spectrum is negligible at small and medium scales (see Fig.\ref{fig:compsep_stats}). We keep in mind that this could potentially introduce bias for cosmological inference applications, and we will address this issue in future work.

\end{document}